\documentclass[twocolumn,prx,aps,amsmath,amssymb,longbibliography]{revtex4-1}

\usepackage{graphicx}
\usepackage{dcolumn}
\usepackage{bm}

\begin{document}

\title{Nonequilibrium finite frequency resonances in differential quantum
  noise driven by Majorana interference}

\author{Sergey Smirnov}
\affiliation{P. N. Lebedev Physical Institute of the Russian Academy of
  Sciences, Department of Solid State Physics, 119991 Moscow, Russia}
\email{1) sergej.physik@gmail.com\\2)
  sergey.smirnov@physik.uni-regensburg.de\\3) ssmirnov@sci.lebedev.ru}

\date{\today}

\begin{abstract}
Nonequilibrium quantum noise $S^>(\omega,V)$ measured at finite frequencies
$\omega$ and bias voltages $V$ probes Majorana bound states in a host
nanostructure via fluctuation fingerprints unavailable in average currents or
static shot noise. When Majorana interference is brought into play, it
enriches nonequilibrium states and makes their nature even more unique. Here
we demonstrate that an interference of two Majorana modes via a nonequilibrium
quantum dot gives rise to a remarkable finite frequency response of the
differential quantum noise $\partial S^>(\omega,V,\Delta\phi)/\partial V$
driven by the Majorana phase difference $\Delta\phi$. Specifically, at low
bias voltages there develops a narrow resonance of width
$\hbar\Delta\omega\sim\sin^2\Delta\phi$ at a finite frequency determined by
$V$, whereas for high bias voltages there arise two antiresonances at two
finite frequencies controlled by both $V$ and $\Delta\phi$. We show that the
maximum and minimum of these resonance and antiresonances have universal
fractional values, $3e^3/4h$ and $-e^3/4h$. Moreover, detecting the
frequencies of the antiresonances provides a potential tool to measure
$\Delta\phi$ in nonequilibrium experiments on Majorana finite frequency
quantum noise.
\end{abstract}

\maketitle

\section{Introduction}\label{intro}
Out of their equilibrium various nanosystems offer a broad spectrum of
measurements in diverse nonequilibrium states achieved nowadays in a wide
range of experiments. Quantum transport provides a versatile tool to
characterize nonequilibrium states via systematically increasing, when
necessary, the depth of complexity to gain a sufficient insight into
microscopic states of a given nanosystem. This is particularly relevant for
nanostructures hosting Majorana bound states (MBSs)
\cite{Kitaev_2001,Alicea_2012,Flensberg_2012,Sato_2016,Aguado_2017,Lutchyn_2018}
designed to perform anyonic fault tolerant quantum computation
\cite{Kitaev_2003} based on non-Abelian manipulations
\cite{Muralidharan_2023}. Although they are supposed to provide an elegant
platform for future quantum computing devices, including implementations of
poor man's MBSs \cite{Tsintzis_2022,Dvir_2023,Tsintzis_2024} in quantum dot
(QD) setups, nonequilibrium response of MBSs is itself an exciting topic rich
of remarkable universal fingerprints. However, not all of these fingerprints
may be uniquely attributed to MBSs. In particular, it is known
\cite{Yu_2021,Frolov_2021} that straightforward measurements of average
electric currents may be unreliable when treating the corresponding
conductances as exclusively induced by MBSs. In this respect quantum transport
experiments measuring the average values of observables may be less
informative than possible thermodynamic approaches where one may uniquely
identify MBSs, for example, by means of the entropy measurements in nanoscopic
systems
\cite{Smirnov_2015,Hartman_2018,Kleeorin_2019,Sela_2019,Smirnov_2021,Smirnov_2021a,
  Pyurbeeva_2021,Ahari_2021,Child_2022,Han_2022,Pyurbeeva_2022,Child_2022a}.
Nevertheless, various average charge and spin currents are extensively
investigated in stationary
\cite{Liu_2011,Fidkowski_2012,Prada_2012,Pientka_2012,Lin_2012,Lee_2013,Kundu_2013,
  Vernek_2014,Ilan_2014,Cheng_2014,Leijnse_2014,Lopez_2014,Lobos_2015,Peng_2015,
  Khim_2015,Sharma_2016,Heck_2016,Das_Sarma_2016,Ramos-Andrade_2016,Lutchyn_2017,
  Weymann_2017,Campo_Jr_2017,Liu_2017,Huang_2017,Liu_2018,Lai_2019,Tang_2020,
  Zhang_2020,Smirnov_2020a,Chi_2021,Wang_2021,He_2021,Galambos_2022,Giuliano_2022,
  Buccheri_2022,Majek_2022,Bondyopadhaya_2022,Zou_2022,Wang_2023,Zou_2023,
  Huguet_2023,Becerra_2023,Ziesen_2023}
and nonstationary \cite{Jin_2022,Zou_2023a,Yao_2023,Wojcik_2024,Taranko_2023}
nonequilibrium Majorana systems and essentially contribute to impressive
progress in characterizing MBSs as much as measurements of average currents
\cite{Mourik_2012,Nadj-Perge_2014,Wang_2022} can in general allow in
contemporary and near future experiments.

As mentioned above, if average electric currents measured in a nanostructure
turn out to be insufficient to uniquely conclude about its properties of
interest, flexibility of quantum transport techniques allows one to
systematically increase the level of complexity and explore, for example,
fluctuations of electric currents to characterize nonequilibrium Majorana
systems beyond fundamental limits imposed by conductance
measurements. Fluctuations of electric currents in nanoscopic systems with
MBSs may be characterized by zero frequency (static) as well as finite
frequency shot noise which has been addressed theoretically
\cite{Liu_2015,Liu_2015a,Haim_2015,Valentini_2016,Zazunov_2016,Smirnov_2017,
  Smirnov_2018,Jonckheere_2019,Manousakis_2020,Feng_2022,Smirnov_2023,Cao_2023}
and recently also in experiments \cite{Ge_2023}. In particular, in QDs
interacting with MBSs, an effective charge is predicted to be fractional, when
MBSs are well separated, or integer, when MBSs strongly overlap and form a
single Dirac fermion \cite{Smirnov_2017}. Another advanced tool to explore
fluctuation fingerprints of MBSs is offered by measurements of quantum noise
$S^>(\omega,V)$ at finite frequencies $\omega$ in a nanosystem whose
nonequilibrium state is maintained by voltages $V$ of, {\it e.g.}, electric or
thermal origin. Here the two finite frequency branches, specifically the
photon emission and absorption noise, may be accessed separately in
nonequilibrium states prepared by a preferred technique, {\it e.g.}, by pure
electric \cite{Bathellier_2019,Smirnov_2019} or thermoelectric
\cite{Smirnov_2019a} means. An important quantity able to reveal universality
of Majorana fluctuations in more detail is the differential quantum noise
$\partial S^>(\omega,V)/\partial V$ having universal units of
$e^3/h$.

Within a nanoscopic device one may create various links between MBSs and
other constituent systems by means of tunneling interactions designed to
crucially involve Majorana tunneling phases to control operation of the whole
device, {\it e.g.}, by means of driven dissipative protocols developed for
Majorana qubits used in quantum computing devices
\cite{Gau_2020,Gau_2020a}. As soon as Majorana tunneling phases are at play,
they may lead to various interference phenomena which give rise to equilibrium
and nonequilibrium characteristics fundamentally absent in systems where MBSs
do not interfere. Therefore, in equilibrium and nonequilibrium nanostructures
hosting MBSs one may expect an extremely unique response in presence of
Majorana interference, especially, if this response is probed via advanced
physical observables such as the entropy and current noise. Indeed, in a
nonequilibrium QD interacting with MBSs via tunneling mechanisms the
differential static shot noise shows a strong dependence on the Majorana
interference and, together with the differential conductance, reveals
universal Majorana fractionalization even when interference effects
significantly suppress both of these observables below their universal unitary
values \cite{Smirnov_2022}. Nevertheless, dynamic fluctuation fingerprints of
Majorana interference cannot be captured by the static shot noise and one has
to resort to other observables such as, {\it e.g.}, finite frequency
differential quantum noise. This observable and its measurements are
particularly attractive because of the following reasons. First, as mentioned
above, similar to the differential conductance, which is measured in universal
units (specifically, in units of $e^2/h$), the differential quantum noise is
also measured in universal units (specifically, in units of $e^3/h$). It is
well known which universal unitary values of the differential conductance
characterize MBSs and these values play the role of a reference for further
research on Majorana mean currents. It is natural to adopt the same strategy
in research on Majorana fluctuation fingerprints and investigate which
universal unitary values associated with resonances or antiresonances of the
differential quantum noise characterize interfering MBSs, especially at finite
frequencies. Second, the differential quantum noise is an experimentally
relevant observable. For example, feasibility of its measurement has been
demonstrated in experiments on Kondo correlated QDs \cite{Basset_2012}. In
particular, quantum detectors allow to separately explore absorption and
emission noise spectra. Third, measurements of only one physical observable,
the differential quantum noise, involve potentially less experimental errors
than, {\it e.g.}, measurements of Fano-like quantities (the differential
quantum noise divided over the differential conductance). Here we would like
to note that the Fano factor is an important physical quantity as proven in
experiments identifying the Laughlin quasiparticle, which is also an exotic
anyon excitation in topological systems emerging due to the fractional quantum
Hall effect \cite{de-Picciotto_1997,Saminadayar_1997}. From a theoretical
point of view, choosing the differential quantum noise or Fano factor should
not make a big difference, because the Fano factor is essentially the
differential quantum noise normalized with the use of the average current. In
practice, however, in strongly nonequilibrium systems, when the
fluctuation-dissipation theorem is broken, fluctuations of a current and its
average value become essentially independent of each other. This requires
measurements of two independent quantities, the differential quantum noise and
differential conductance. Thus errors from experimental measurements of the
two physical observables may accumulate and result in less precise
outcomes. Fourth, the differential quantum noise at finite frequencies might
be helpful in understanding of whether a bound state in the continuum (BIC)
has a Majorana origin. That could be relevant because mean currents might be
controversial in this respect. Below, as an alternative view, we additionally
propose an interpretation of our results on the finite frequency differential
quantum noise in terms of the Majorana BIC discussed previously only within
the differential conductance \cite{Ramos-Andrade_2019}.

In this paper we numerically investigate the differential quantum noise
$\partial S^>(\omega,V,\Delta\phi)/\partial V$ at finite frequencies $\omega$
when it is driven by interference of MBSs linked to a QD via tunneling
interactions. Nonequilibrium states of the QD are induced by a bias voltage
$V$ and controlled by the difference $\Delta\phi$ of the Majorana tunneling
phases. We consider regimes of both low and high bias voltages $V$. First, for
small values of $V$ it is shown that the differential quantum noise
$\partial S^>(\omega,V,\Delta\phi)/\partial V$ as a function of the frequency
$\omega$ has a step-like shape when the Majorana interference is absent. The
height of this Majorana step is equal to the universal unitary value
$e^3/4h$. As soon as the Majorana interference emerges, there develops a
finite frequency resonance on top of the Majorana step at a frequency whose
value is specified by $V$. The width of this resonance is proportional to
$\sin^2\Delta\phi$ and its maximum is given by the universal unitary value
$3e^3/4h$. Second, for large values of $V$ we demonstrate that in absence of
the Majorana interference $\partial S^>(\omega,V,\Delta\phi)/\partial V$ is
strongly suppressed at all frequencies except for a vicinity of a finite
frequency specified by $V$ where it exhibits an antiresonance with the
universal unitary minimum $-e^3/4h$. When the Majorana interference is
switched on, this antiresonance is split into two antiresonances having the
same universal unitary minimum $-e^3/4h$ located at two finite frequencies
specified by both $V$ and $\Delta\phi$.

The paper is organized as follows. In Section \ref{mnqdmi} we present a model
of a nonequilibrium QD coupled to MBSs which may interfere on the QD for
finite values of the Majorana phase difference. Nonequilibrium behavior of
this system may properly be described using the Keldysh technique which is
applied within the formalism of the Keldysh field integral in Section
\ref{qnkfi}. We demonstrate and discuss the numerical results obtained for the
finite frequency differential quantum noise in Section \ref{nrdqnff}. Finally,
Section \ref{concl} provides conclusions and outlooks.
\section{Model of a nonequilibrium quantum dot with Majorana interference}\label{mnqdmi}
A minimal platform where one may access fluctuation fingerprints of Majorana
interference in finite frequency quantum noise is provided by a QD with one
nondegenerate state interacting via tunneling with a grounded topological
superconductor (TS) whose low energy behavior is governed by two MBSs located
at its ends.

The QD Hamiltonian has the form
\begin{equation}
  \hat{H}_\text{QD}=\epsilon_d d^\dagger d,
  \label{Ham_QD}
\end{equation}
where $\epsilon_d$ is the single-particle energy of the QD state. The position
of the energy level $\epsilon_d$ with respect to the chemical potential $\mu$
of the system may be controlled by a gate voltage. As stated in
Ref. \cite{Flensberg_2011}, the physical reason to consider a nondegenerate QD
is that the topological superconducting phase requires high magnetic fields
removing the spin degeneracy in the QD and thus only one spin component is of
relevance. Numerical renormalization group calculations \cite{Tijerina_2015},
showing the linear conductance plateau $e^2/2h$ in high magnetic fields,
provide an exact support for that statement. As a consequence, Kondo
correlations have no impact on the Majorana induced behavior and a
noninteracting QD is a proper model. Below we will focus on the situation when
$\epsilon_d>0$ meaning that the QD is empty. When low energy dynamics is
dominated by MBSs, it plays no role whether $\epsilon_d<0$ or $\epsilon_d>0$
because dependence on the gate voltage, {\it i.e.} on $\epsilon_d$, becomes
very weak due to the Majorana universality. However, we prefer using positive
values of $\epsilon_d$ for consistency with possible future experiments where
for $\epsilon_d<0$ there could remain some residual Kondo correlations
inducing the universal Kondo behavior (see
Refs. \cite{Hewson_1997,Smirnov_2011a,Niklas_2016}) of physical observables
despite high magnetic fields. The Kondo universality present for
$\epsilon_d<0$ would definitely be eliminated for $\epsilon_d>0$ restricting
such experiments to a regime where one may safely observe the Majorana
universality.
\begin{figure}
\includegraphics[width=8.0 cm]{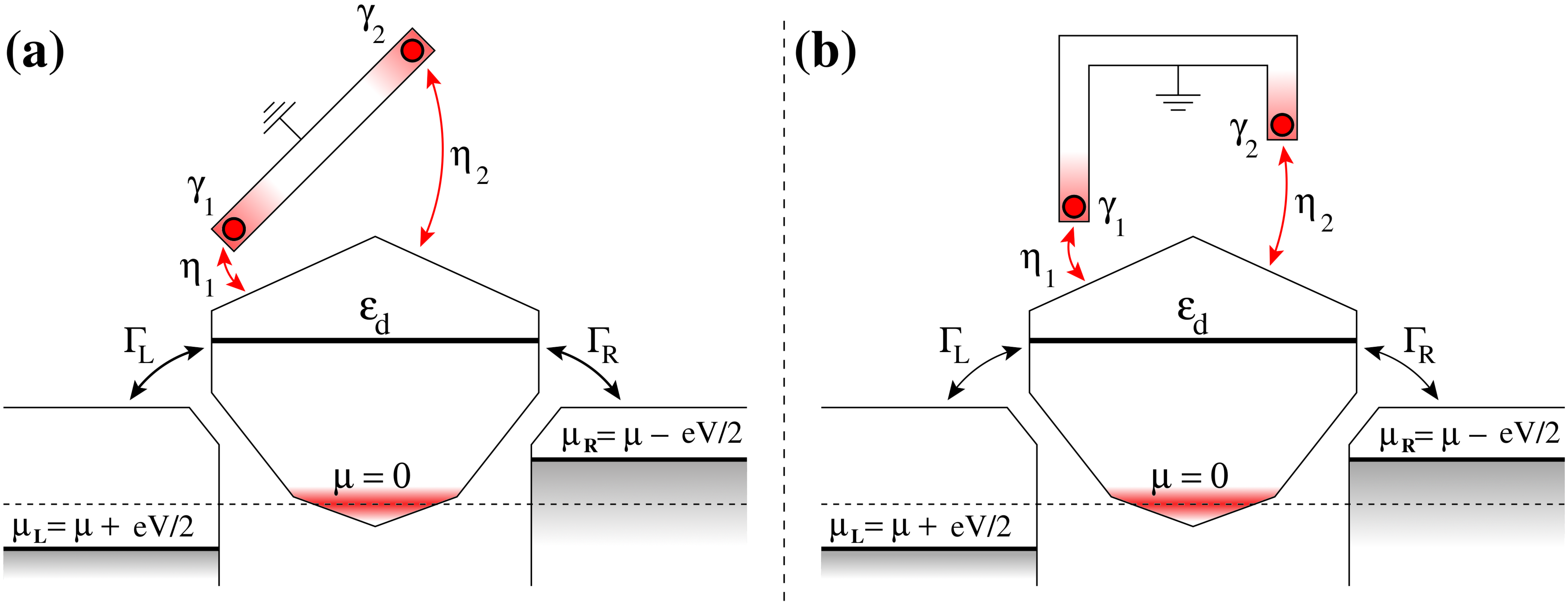}
\caption{\label{figure_1} A schematic outline of the mathematical model
  formulated in Eqs. (\ref{Ham_QD})-(\ref{Ham_QD-C}), assuming $eV<0$. The
  model may be   realized, for example, with a straight TS, as shown in ({\bf
    a}), or with an   arched TS, as shown in ({\bf b}).}
\end{figure}

The Hamiltonian of the TS is
\begin{equation}
  \hat{H}_\text{TS}=\frac{i}{2}\xi\gamma_2\gamma_1,
  \label{Ham_TS}
\end{equation}
where the Majorana operators satisfy $\gamma_k^\dagger=\gamma_k$,
$\{\gamma_k,\gamma_j\}=2\delta_{kj}$, $k,j=1,2$ and a finite overlap of the
MBSs is taken into account by finite values of the energy $\xi$. The
interaction between the QD and TS is described by the following tunneling
Hamiltonian
\begin{equation}
  \hat{H}_\text{QD-TS}=\eta_1^*d^\dagger\gamma_1+\eta_2^*d^\dagger\gamma_2+\text{H.c.}
  \label{Ham_QD-TS}
\end{equation}
Here the tunneling matrix elements have the form
\begin{equation}
  \eta_k=|\eta_k|e^{i\phi_k},\quad k=1,2.
  \label{QD-TS_mtrx_el}
\end{equation}
The amplitude $|\eta_k|$ describes the strength of the tunneling coupling
between the QD and Majorana mode $\gamma_k$, $k=1,2$. Below we will assume a
specific preparation of the system with $|\eta_1|\gg|\eta_2|$. This
corresponds to a relatively simpler technological preparation of the QD
located at unequal distances to the ends of the TS as, for example, in
Ref. \cite{Deng_2018}. Specifically, the QD is located closer to the Majorana
mode $\gamma_1$. Implementations with $|\eta_1|\gtrsim |\eta_2|$, that is with
a more symmetric location of the QD with respect to the ends of the TS, would
require an advanced technology, for example, preparing a TS with a more
complex shape \cite{Liu_2011}. The Majorana tunneling phases $\phi_k$,
$k=1,2$, are of particular importance. The tunneling phase difference
$\Delta\phi=\phi_1-\phi_2$ gives rise to Majorana interference in the
system. Various physical observables acquire a dependence on $\Delta\phi$. In
particular, fluctuation fingerprints of the Majorana interference in the
finite frequency quantum noise are encoded in its dependence on the Majorana
phase difference $\Delta\phi$.

The electric currents, in particular their fluctuations, are measured in two
normal metallic contacts coupled via tunneling to the QD. The two contacts are
denoted as left ($L$) and right ($R$). Their Hamiltonian is
\begin{equation}
  \hat{H}_\text{C}=\sum_{l=L,R}\sum_k\epsilon_kc_{lk}^\dagger c_{lk},
  \label{Ham_C}
\end{equation}
In Eq. (\ref{Ham_C}) it is assumed that both contacts have the same continuum
energy spectrum $\epsilon_k$. It gives rise to a certain density of states
$\nu(\epsilon)$ which is a function of energy. The continuum energy spectrum
of the contacts is involved in various physical observables only through the
density of states whose energy dependence is often neglected,
$\nu(\epsilon)\approx\nu_C/2$. This assumes a weak energy dependence of
$\nu(\epsilon)$ in the energy domain where quantum transport is most effective
\cite{Altland_2010}. The contacts are in their equilibrium states which are
characterized by the corresponding Fermi-Dirac distributions,
\begin{equation}
  n_{L,R}(\epsilon)=\frac{1}{\exp\bigl(\frac{\epsilon-\mu_{L,R}}{k_\text{B}T}\bigr)+1}.
  \label{F-D_distrib_L_R}
\end{equation}
In Eq. (\ref{F-D_distrib_L_R}) we assume that the contacts have the same
temperature $T$ but their chemical potentials,
\begin{equation}
  \mu_{L,R}=\mu\pm eV/2,
  \label{Chem_pot_L_R}
\end{equation}
are different for finite bias voltages $V$. Below, in Section \ref{nrdqnff},
we perform numerical calculations assuming that the bias voltage is chosen
such that $eV<0$.

The tunneling Hamiltonian taking into account the coupling between the
contacts and QD has the form
\begin{equation}
  \hat{H}_\text{QD-C}=\sum_{l=L,R}\mathcal{T}_l\sum_kc_{lk}^\dagger d+\text{H.c.},
  \label{Ham_QD-C}
\end{equation}
assuming independence of the tunneling matrix elements of the quantum numbers
$k$ characterizing the states of the contacts. The matrix elements
$\mathcal{T}_{L,R}$ together with the density of states $\nu_C$ of the
contacts determine the strengths $\Gamma_{L,R}=\pi\nu_C|\mathcal{T}_{L,R}|^2$
of the tunneling interactions between the QD and, respectively, the left and
right contacts. Below we focus on the situation when $\Gamma_L=\Gamma_R$. The
total tunneling strength is $\Gamma\equiv\Gamma_L+\Gamma_R$.

The system described by Eqs. (\ref{Ham_QD})-(\ref{Ham_QD-C}) is schematically
illustrated in Fig. \ref{figure_1}. It may be implemented using various
technological structures. For example, the straight TS in
Fig. \ref{figure_1}(a) corresponds to the structure proposed in
Ref. \cite{Deng_2018} while the arched TS in Fig. \ref{figure_1}(b)
corresponds to the structure proposed in Ref. \cite{Flensberg_2011} and used,
{\it e.g.}, in Refs. \cite{Liu_2011,Ricco_2018}.
\section{Quantum noise from the Keldysh field integral}\label{qnkfi}
When $V\neq 0$, the system is brought into a nonequilibrium state which has to
be dealt with by a proper theoretical tool. Here we use the Keldysh technique
implemented within the formalism of the Keldysh field integral
\cite{Altland_2010} defined on the Keldysh closed time contour $C_\text{K}$
with forward and backward branches labeled, respectively, with $q=\pm$. The
general strategy of this formalism assumes a proper formulation of a fermionic
problem in terms of coherent states and their eigenvalues which are the
Grassmann fields $\chi(t)$, $t\in C_\text{K}$, corresponding to the
annihilation and creation operators used to express the Hamiltonian of this
problem.

Our problem is described by the Hamiltonian
\begin{equation}
  \hat{H}=\hat{H}_\text{QD}+\hat{H}_\text{TS}+\hat{H}_\text{QD-TS}+\hat{H}_C+\hat{H}_\text{QD-C}.
  \label{Ham}
\end{equation}
In accordance with the general strategy of the Keldysh field integral, we
introduce the Grassmann fields $(\psi(t),\bar{\psi}(t))$,
$(\phi_{lk}(t),\bar{\phi}_{lk}(t))$ and $(\zeta(t),\bar{\zeta}(t))$ instead of
the operators $(d,d^\dagger)$, $(c_{lk},c_{lk}^\dagger)$ and
$(\gamma_1,\gamma_2)$. Here the bars over the Grassmann fields denote the
Grassmann conjugation (G.c.). Various physical observables of interest may be
expressed via these Grassmann fields. In particular, the matrix element of the
electric current operator between proper coherent states is
\begin{equation}
  I_{lq}(t)=\frac{ie}{\hbar}\sum_k\bigl[\mathcal{T}_l\bar{\phi}_{lkq}(t)\psi_q(t)-\text{G.c.}\bigr],
  \label{El_current_mtrx_el}
\end{equation}
where $l=L,R$ specifies the contact, $q=\pm$ the branch of $C_\text{K}$ and
$t$ real time.

The electric current and other physical quantities may be obtained from the
Keldysh generating functional which is a field integral over the fields
defined on $C_\text{K}$
\begin{equation}
  [\bar{\Phi}(t),\Phi(t)]\equiv[\bar{\psi}(t),\psi(t);\bar{\phi}_{lk}(t),\phi_{lk}(t);\bar{\zeta}(t),\zeta(t)]
  \label{Composite_fields}
\end{equation}
with the integrand specified by the Keldysh action $S_\text{K}$,
\begin{equation}
  Z[J_{lq}(t)]=\int\mathcal{D}[\bar{\Phi}(t),\Phi(t)]e^{\frac{i}{\hbar}S_\text{K}[J_{lq}(t)]},
  \label{Keld_gf}
\end{equation}
where the Keldysh action is the sum of the actions of the QD, TS, contacts,
tunneling actions and a source action to generate physical quantities of
interest,
\begin{equation}
  \begin{split}
    &S_\text{K}[J_{lq}(t)]=S_\text{QD}[\bar{\psi}_q(t),\psi_q(t)]\\
    &+S_\text{TS}[\bar{\zeta}_q(t),\zeta_q(t)]+S_\text{C}[\bar{\phi}_{lkq}(t),\phi_{lkq}(t)]\\
    &+S_\text{QD-TS}[\bar{\psi}_q(t),\bar{\zeta}_q(t);\psi_q(t),\zeta_q(t)]\\
    &+S_\text{QD-C}[\bar{\psi}_q(t),\bar{\phi}_{lkq}(t);\psi_q(t),\phi_{lkq}(t)]\\
    &+S_\text{SRC}[J_{lq}(t)].
    \end{split}
  \label{Keld_act}
\end{equation}
In Eq. (\ref{Keld_act}) the actions $S_\text{QD}$, $S_\text{TS}$ and
$S_\text{C}$ are represented by standard $2\times 2$ matrices in the
retarded-advanced space (see Ref. \cite{Altland_2010}). The form of the
tunneling actions is obtained from the Hamiltonians in Eqs. (\ref{Ham_QD-TS})
and (\ref{Ham_QD-C}):
\begin{equation}
  \begin{split}
    &S_\text{QD-TS}[\bar{\psi}_q(t),\bar{\zeta}_q(t);\psi_q(t),\zeta_q(t)]=\\
    &-\int_{-\infty}^\infty dt \{\eta_1^*[\bar{\psi}_+(t)\zeta_+(t)+\bar{\psi}_+(t)\bar{\zeta}_+(t)\\
    &-\bar{\psi}_-(t)\zeta_-(t)-\bar{\psi}_-(t)\bar{\zeta}_-(t)]+i\eta_2^*[\bar{\psi}_+(t)\zeta_+(t)\\
    &+\bar{\psi}_-(t)\bar{\zeta}_-(t)-\bar{\psi}_-(t)\zeta_-(t)-\bar{\psi}_+(t)\bar{\zeta}_+(t)]+\text{G.c.}\}
  \end{split}
  \label{Act_QD-TS}
\end{equation}
\begin{equation}
  \begin{split}
    &S_\text{QD-C}[\bar{\psi}_q(t),\bar{\phi}_{lkq}(t);\psi_q(t),\phi_{lkq}(t)]=\\
    &-\int_{-\infty}^\infty dt\sum_{l=L,R}\sum_k\{\mathcal{T}_l[\bar{\phi}_{lk+}(t)\psi_+(t)\\
    &-\bar{\phi}_{lk-}(t)\psi_-(t)]+\text{G.c.}\}.
  \end{split}
  \label{Act_QD-C}
\end{equation}
To explore the electric current the source action is chosen as follows:
\begin{equation}
  S_\text{SRC}[J_{lq}(t)]=-\int_{-\infty}^\infty dt\sum_{l=L,R}\sum_{q=\pm}J_{lq}(t)I_{lq}(t).
  \label{Act_SRC}
\end{equation}
From the Keldysh generating functional one obtains the average electric
current and current-current correlations by differentiations of $Z[J_{lq}(t)]$
over the source field $J_{lq}(t)$:
\begin{equation}
  \langle I_{lq}(t)\rangle_0=i\hbar\frac{\delta Z[J_{lq}(t)]}{\delta J_{lq}(t)}\biggl|_{J_{lq}(t)=0}
  \label{Avr_el_current}
\end{equation}
\begin{equation}
  \langle I_{lq}(t)I_{l'q'}(t')\rangle_0=(i\hbar)^2\frac{\delta^2 Z[J_{lq}(t)]}{\delta J_{lq}(t)\delta J_{l'q'}(t')}\biggl|_{J_{lq}(t)=0}.
  \label{Current-current_corr}
\end{equation}
In Eqs. (\ref{Avr_el_current}) and (\ref{Current-current_corr}) averaging,
\begin{equation}
  \begin{split}
    &\langle \mathcal{F}[\bar{\Phi}(t_j),\Phi(t_{j'})]\rangle_0\equiv\\
    &\int\mathcal{D}[\bar{\Phi}(t),\Phi(t)]e^{\frac{i}{\hbar}S_\text{K}^{(0)}}\mathcal{F}[\bar{\Phi}(t_j),\Phi(t_{j'})]
  \end{split}
  \label{Avr_zero_src}
\end{equation}
is performed using the Keldysh action without the sources,
$S_\text{K}^{(0)}\equiv S_\text{K}[J_{lq}(t)=0]$.

For measurements of the electric current and its fluctuations in the left
contact we put below $l=L$. The mean electric current $I(V,\Delta\phi)$ as a
function of the bias voltage $V$ and the Majorana tunneling phase difference
$\Delta\phi$ is obtained from Eq. (\ref{Avr_el_current}) with $l=L$, that is
$I(V,\Delta\phi)=\langle I_{Lq}(t)\rangle_0$, where $q$ is arbitrary since the
average $\langle\cdots\rangle_0$ defined in Eq. (\ref{Avr_zero_src}) removes
any dependence on $q$. We are interested in the quantum noise which is defined
as the greater current-current correlator,
\begin{equation}
  \begin{split}
  &S^>(t,t';V,\Delta\phi)\equiv\langle\delta I_{L-}(t)\delta I_{L+}(t')\rangle_0,\\
  &\delta I_{Lq}(t)\equiv I_{Lq}(t)-I(V,\Delta\phi).
  \end{split}
  \label{Quant_noise_time}
\end{equation}
Since we deal with stationary nonequilibrium states, the quantum noise depends
on $t$ and $t'$ only through their difference,
$S^>(t,t';V,\Delta\phi)=S^>(t-t';V,\Delta\phi)$. The Fourier transform
\begin{equation}
  S^>(\omega,V,\Delta\phi)=\int_{-\infty}^\infty dt e^{i\omega t}S^>(t;V,\Delta\phi)
  \label{Quant_noise_ff}
\end{equation}
provides the finite frequency quantum noise which, depending on the sign of
the frequency $\omega$, probes the photon absorption/emission spectra:
\begin{equation}
  S^{\text{ab}/\text{em}}(\omega,V,\Delta\phi)=
  \begin{cases}
    S^>(\omega,V,\Delta\phi),&\quad \omega>0\\
    S^>(\omega,V,\Delta\phi),&\quad \omega<0.
  \end{cases}
  \label{Photon_ab_em_spectra}
\end{equation}
In our numerical calculations we have only focused on positive frequencies
which admit interpretations of the numerical results in terms of photon
absorption processes (see Section \ref{nrdqnff} for details).

To examine universal fingerprints of the fluctuation behavior induced by the
interference of the MBSs we numerically calculate the differential quantum
noise, $\partial S^>(\omega,V,\Delta\phi)/\partial V$, which is an
experimentally accessible physical quantity \cite{Basset_2012}. In the
numerical calculations we first obtain the finite frequency quantum noise
$S^>(\omega,V,\Delta\phi)$ from numerical integrations with high
precision. Afterwards, using finite differences, we numerically calculate the
derivatives $\partial S^>(\omega,V,\Delta\phi)/\partial V$ at finite
frequencies $\omega$ in a wide rage. Of particular interest is the universal
Majorana regime arising when the energy scale characterizing the strength of
the tunneling between the QD and TS essentially exceeds the other energy
scales of the problem. It may be achieved, for example, if the parameters
satisfy the following inequality
\begin{equation}
  |\eta_1|>\text{max}\{|\epsilon_d|,k_\text{B}T,|\eta_2|,\xi,\Gamma,|eV|\},
  \label{U_M_regime}
\end{equation}
which has been assumed in performing our numerical analysis of the finite
frequency differential quantum noise discussed in the next section. To
interpret the behavior of the differential quantum noise at finite frequencies
in terms of emission and absorption processes it may be helpful to recall (see
Ref. \cite{Smirnov_2019}) that in the parameter regime specified by
Eq. (\ref{U_M_regime}) the MBSs induce in the QD a quasiparticle zero-energy
state as well as quasiparticle states with the energies $\mp 2|\eta_1|$. Due
to the coupling of the QD to the contacts the width of the zero-energy state
is equal to $\Gamma$ whereas the widths of the states with the energies $\mp
2|\eta_1|$ are both equal to $\Gamma/2$. The behavior of the differential
quantum noise discussed below is mainly governed by the zero-energy
quasiparticle state. Transport processes with initial and final states in the
contacts or TS may excite the QD. In the next section under weak excitations
of the QD we understand those excitations which are localized within the
energy range of order $\Gamma$ around the zero-energy state. Although, as
discussed in the next section, large excitations to the states with the
energies $\mp 2|\eta_1|$ also occur, they are not in the focus of the present
work.
\begin{figure}
\includegraphics[width=8.0 cm]{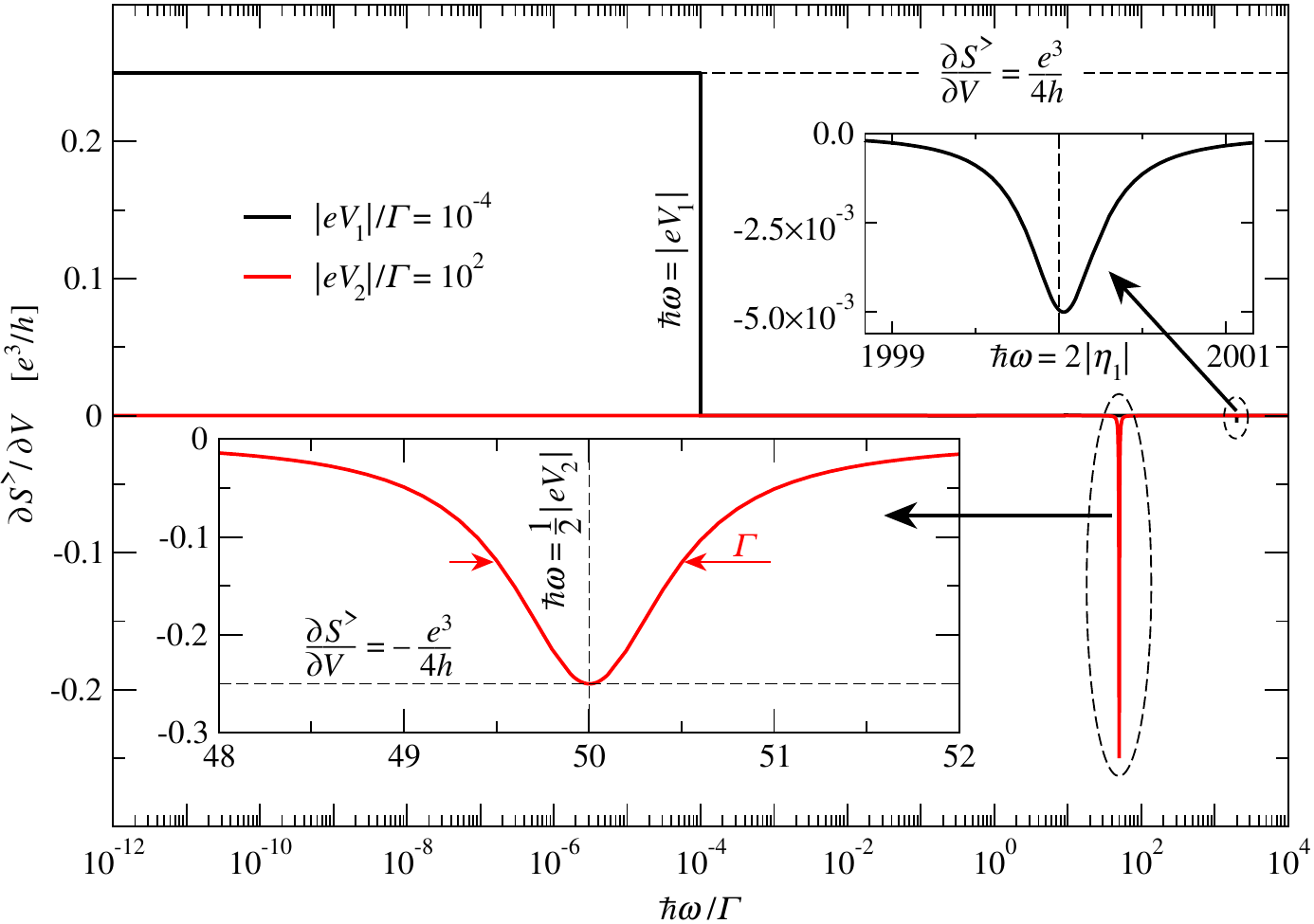}
\caption{\label{figure_2} Differential quantum noise
  $\partial S^>(\omega,V)/\partial V$ as a function of the frequency $\omega$
  in the absence of direct tunneling between the QD and the second Majorana
  mode $\gamma_2$, that is $|\eta_2|=0$. The black curve corresponds to a low
  bias voltage, $V=V_1$, whereas the red curve corresponds to a high bias
  voltage, $V=V_2$, with the specific values $|eV_1|/\Gamma=10^{-4}$ and
  $|eV_2|/\Gamma=10^2$. Here $\epsilon_d/\Gamma=10$, $k_BT/\Gamma=10^{-12}$,
  $|\eta_1|/\Gamma=10^3$, $\xi/\Gamma=10^{-14}$.}
\end{figure}
\section{Numerical results for the differential quantum noise at finite frequencies}\label{nrdqnff}
We start our numerical analysis with the situation when Majorana interference
is absent, that is, when there is no any dependence on the Majorana tunneling
phase difference $\Delta\phi$. This happens for $|\eta_2|=0$. In this case, as
can be seen in Fig. \ref{figure_2}, at low bias voltages, $|eV|\ll\Gamma$, the
differential quantum noise as a function of the frequency $\omega$ has a
step-like shape (the black curve) with the jump located at
$\hbar\omega=|eV|$. At frequencies $\hbar\omega<|eV|$ the differential quantum
noise does not depend on $\omega$ forming a plateau with the universal unitary
value $e^3/4h$ known for the static limit, $\omega\rightarrow 0$ (see,
{\it e.g.}, Refs. \cite{Liu_2015} and \cite{Smirnov_2017}). At frequencies
$\hbar\omega>|eV|$ the differential quantum noise is strongly suppressed
except for a vicinity of the frequency $\hbar\omega=2|\eta_1|$. Since at
positive frequencies the quantum noise may be interpreted in terms of photon
absorption processes (see Section \ref{qnkfi},
Eq. (\ref{Photon_ab_em_spectra}), and also
Refs. \cite{Bathellier_2019,Smirnov_2019}), such a behavior at low bias
voltages indicates that for a given value $V$ of the bias voltage,
$|eV|\ll\Gamma$, photon absorption becomes impossible for $\hbar\omega>|eV|$
except for a vicinity of $\hbar\omega=2|\eta_1|$ where a photon absorption
\begin{figure}
\includegraphics[width=8.0 cm]{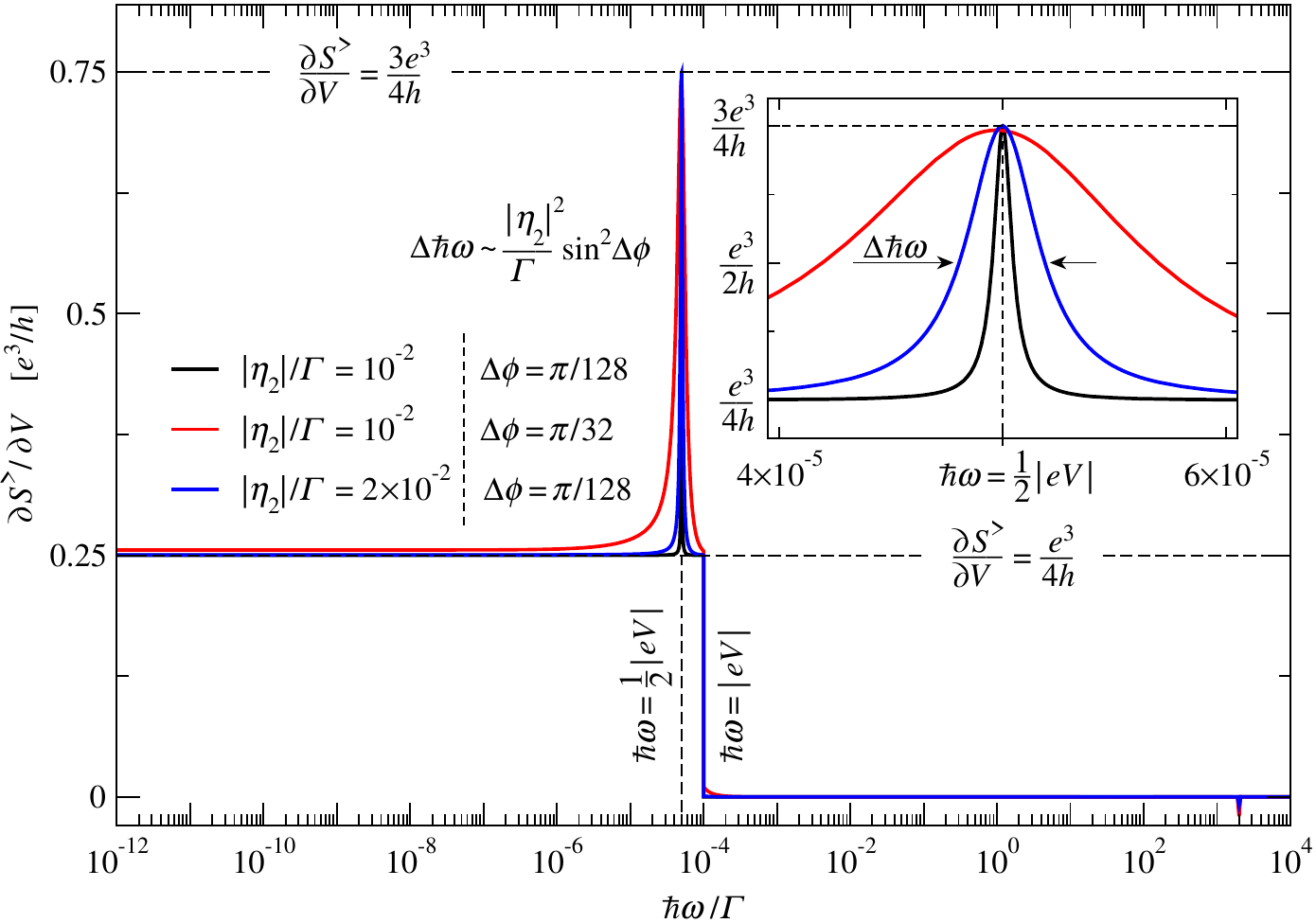}
\caption{\label{figure_3} Differential quantum noise
  $\partial S^>(\omega,V,\Delta\phi)/\partial V$ as a function of the
  frequency $\omega$ at low bias voltages, $|eV|\ll\Gamma$, and in the
  presence of direct tunneling between the QD and the second Majorana mode
  $\gamma_2$. For all the curves $|eV|/\Gamma=10^{-4}$. The black curve:
  $|\eta_2|/\Gamma=10^{-2}$, $\Delta\phi=\pi/128$. The red curve:
  $|\eta_2|/\Gamma=10^{-2}$, $\Delta\phi=\pi/32$. The blue curve:
  $|\eta_2|/\Gamma=2\times 10^{-2}$, $\Delta\phi=\pi/128$. The other
  parameters are the same as in Fig. \ref{figure_2}.}
\end{figure}
becomes possible due to transport processes occurring along with the
excitation of the QD by energy $\Delta\epsilon=2|\eta_1|$ (see
Ref. \cite{Smirnov_2019}). The upper inset shows a shallow antiresonance
corresponding to such photon absorption. It is located around
$\hbar\omega=2|\eta_1|$ and the full width of this antiresonance at half of
its minimum is equal to $\Gamma/2$. It is interesting to note that at low bias
voltages, $|eV|\ll\Gamma$, photon absorption admitted by tunneling of
quasiparticles from the left contact to the TS does not contribute to the
differential quantum noise $\partial S^>(\omega,V)/\partial V$. Indeed, when
accompanied by weak excitations of the QD, such tunneling processes would
increase the quasiparticle energy by $\Delta\epsilon_{qp}=|eV|/2$ and thus in
a vicinity of the frequency $\hbar\omega=|eV|/2$ one could expect that
$\partial S^>(\omega,V)/\partial V$ has a specific resonance or
antiresonance. This is, however, not the case as demonstrated by the black
curve. On the other side, the situation at high bias voltages is qualitatively
different. As demonstrated by the red curve, the differential quantum noise
$\partial S^>(\omega,V)/\partial V$ is suppressed by high bias voltages,
$|eV|\gg\Gamma$, at all frequencies except for a vicinity of
$\hbar\omega=|eV|/2$. Around this frequency there develops an antiresonance in
$\partial S^>(\omega,V)/\partial V$. As shown in the lower inset, the minimum
of this antiresonance is located at $\hbar\omega=|eV|/2$ where the
differential quantum noise reaches the universal unitary value $-e^3/4h$. The
full width $\Delta\hbar\omega$ of the antiresonance at half of its minimum is
equal to $\Gamma$,
\begin{equation}
  \begin{split}
    &\frac{\partial S^>(\omega,V)}{\partial V}\biggl|_{\hbar\omega=|eV|/2}=-\frac{e^3}{4h},\\
    &\Delta\hbar\omega=\Gamma.
  \end{split}
  \label{Uumin_hbv}
\end{equation}
Such a frequency dependence of $\partial S^>(\omega,V)/\partial V$
reveals that at large bias voltages, $|eV|\gg\Gamma$, the most efficient
opening of a photon-absorption channel occurs in a neighborhood of
$\hbar\omega=|eV|/2$ where it is admitted by quasiparticle tunneling from the
left contact to the TS. These tunneling processes occur together with weak
excitations of the QD in such a way that in the final state the quasiparticle
energy is increased by $\Delta\epsilon_{qp}=|eV|/2$ and the QD energy is
increased or decreased by $\Delta\epsilon\lesssim\Gamma$ in accordance with
the location of the minimum and the characteristic width of the observed
antiresonance.

\begin{figure}
\includegraphics[width=8.0 cm]{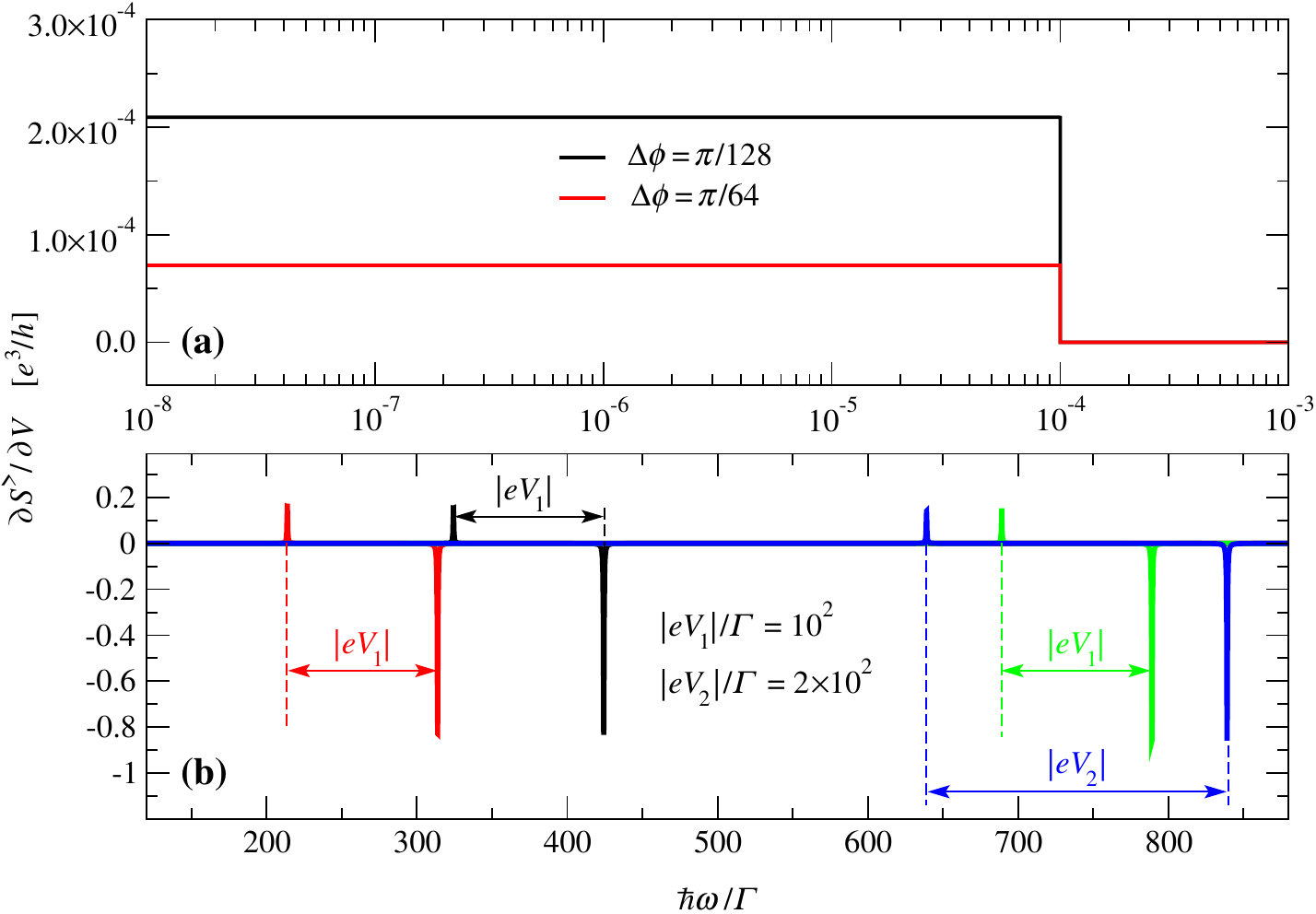}
\caption{\label{figure_8} Differential quantum noise
  $\partial S^>(\omega,V,\Delta\phi)/\partial V$ as a function of the
  frequency $\omega$ for ABSs emerging in our model when the Majorana
  tunneling amplitudes have the same order, $|\eta_2|\sim |\eta_1|$, and the
  overlap energy $\xi$ is large. Specifically, here we use
  $\xi/\Gamma=8\times 10^2$. Panel ({\bf a}): Low bias voltages,
  $|eV|\ll\Gamma$, with the specific values of the parameters
  $|\eta_2|/\Gamma=2\times 10^2$, $|eV|/\Gamma=10^{-4}$ and
  $\Delta\phi=\pi/128$ (black curve), $\Delta\phi=\pi/64$ (red curve).
  Panel ({\bf b}): High bias voltages, $|eV|\gg\Gamma$. The black curve:
  $|\eta_2|/\Gamma=2\times 10^2$, $\Delta\phi=\pi/2$, $|eV|=|eV_1|$. The red
  curve: $|\eta_2|/\Gamma=2\times 10^2$, $\Delta\phi=\pi/4$,
  $|eV|=|eV_1|$. The green curve: $|\eta_2|/\Gamma=4\times 10^2$,
  $\Delta\phi=\pi/2$, $|eV|=|eV_1|$. The blue curve:
  $|\eta_2|/\Gamma=4\times 10^2$, $\Delta\phi=\pi/2$, $|eV|=|eV_2|$. Here
  $|eV_1|/\Gamma=10^2$, $|eV_2|/\Gamma=2\times 10^2$. The other parameters are
  the same as in Fig. \ref{figure_2}.}
\end{figure}

When $|\eta_2|$ is finite, it provides a direct tunneling mechanism between
the QD and the second Majorana mode $\gamma_2$. Now the first and second
Majorana modes, $\gamma_1$ and $\gamma_2$, meet at the QD where they may
interfere. This interference is controlled by the difference of the Majorana
tunneling phases $\Delta\phi=\phi_1-\phi_2$ which essentially determines the
differential quantum noise and brings fundamental changes in its frequency
dependence. As one can see in Fig. \ref{figure_3}, in contrast to the case
with $|\eta_2|=0$ and $|eV|\ll\Gamma$, here for all the three curves, in
addition to the already known step-like shape, one observes a resonance in
$\partial S^>(\omega,V,\Delta\phi)/\partial V$ in a vicinity of the frequency
$\hbar\omega=|eV|/2$. As in Fig. \ref{figure_2}, on the plateau of the step
$\partial S^>(\omega,V,\Delta\phi)/\partial V=e^3/4h$ whereas the maximum of
the resonance arising on top of this plateau reaches another universal unitary
value,
\begin{equation}
  \frac{\partial S^>(\omega,V,\Delta\phi)}{\partial V}\biggl|_{\hbar\omega=|eV|/2}=\frac{3e^3}{4h}.
  \label{Uumax_lbv}
\end{equation}
The inset shows in more detail the shape of the resonance in each of the three
curves, particularly, the variation of its full width $\Delta\hbar\omega$ at
half of its maximum. It is clearly seen that $\Delta\hbar\omega$ strongly
depends on both $|\eta_2|$ and $\Delta\phi$. Our numerical calculations reveal
that
\begin{equation}
  \Delta\hbar\omega\sim \frac{(|\eta_2|\sin\Delta\phi)^2}{\Gamma}.
  \label{FWHM_res_lbv}
\end{equation}
In accordance with Fig. \ref{figure_2}, the resonance disappears for
$|\eta_2|=0$. Moreover, it also disappears when $\Delta\phi=0,\pi$. The strong
dependence of this resonance on the Majorana phase difference $\Delta\phi$ is
suggestive of its pure Majorana interference nature. It also may be
interpreted as a bound state in the continuum related to the zero-energy peak
in the spectral function of the TS (see, {\it e.g.},
Ref. \cite{Ramos-Andrade_2019}). The Majorana interference projects this
zero-energy peak in the spectral function onto the above resonance of finite
width $\Delta\hbar\omega$ in the differential quantum noise and in this way
turns it into a quasi-BIC. This BIC may also be revealed by means of
interference effects captured by the conductance but that would require a more
complicated system with at least two TSs where the BIC manifests as a
zero-bias antiresonance in the conductance as has been shown in
Ref. \cite{Ramos-Andrade_2019}. In contrast, the differential quantum noise at
finite frequencies enables one to detect this BIC in a simpler system using
only one TS. Here at low bias voltages, $|eV|\ll\Gamma$, photon absorption
processes admitted by weak excitations of the QD and the quasiparticle
tunneling from the left contact to the TS are activated by the Majorana
interference and reveal the BIC in a vicinity of the frequency
$\hbar\omega=|eV|/2$ as the above discussed resonance with the universal
unitary maximum $3e^3/4h$. More importantly, since Majorana fluctuation
fingerprints are more unique than those of Majorana mean currents, the
differential quantum noise at finite frequencies provides a reliable tool to
identify BICs as originating from Majorana states and not from other states
having non-Majorana nature. In particular, as will be shown below (see
Fig. \ref{figure_8} and the corresponding discussion), the differential
quantum noise does not exhibit any resonance at the frequency
$\hbar\omega=|eV|/2$ when the MBSs are replaced with Andreev bound states
(ABSs) coupled to the QD. Thus, the resonance in
\begin{figure}
\includegraphics[width=8.0 cm]{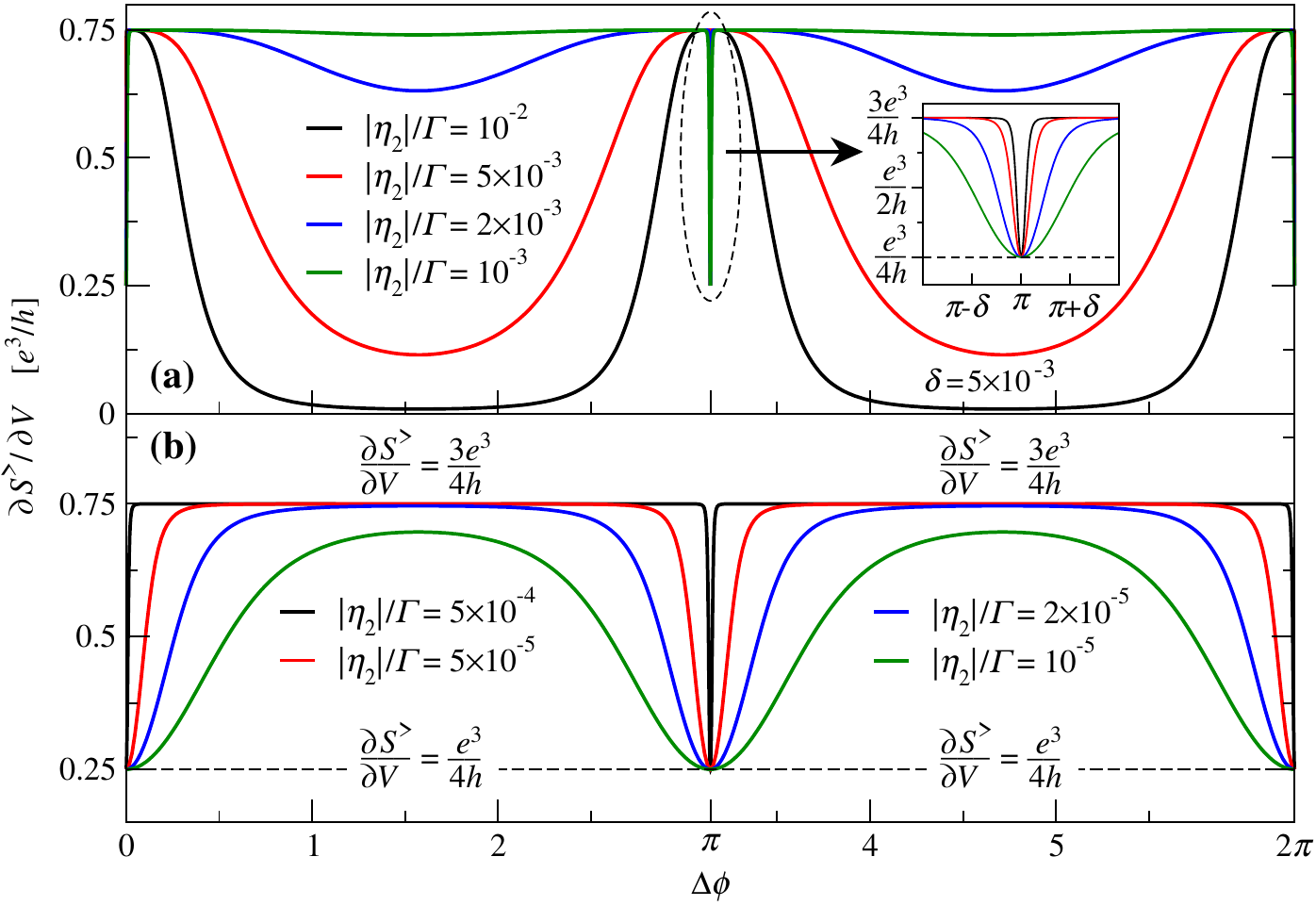}
\caption{\label{figure_4} Differential quantum noise
  $\partial S^>(\omega,V,\Delta\phi)/\partial V$ as a function of the Majorana
  phase difference $\Delta\phi$ at $\hbar\omega=|eV|/2$,
  $|eV|/\Gamma=10^{-4}$. Panel ({\bf a}): $|\eta_2|/\Gamma=10^{-2}$ (black
  curve), $|\eta_2|/\Gamma=5\times 10^{-3}$ (red curve),
  $|\eta_2|/\Gamma=2\times 10^{-3}$ (blue curve), $|\eta_2|/\Gamma=10^{-3}$
  (green curve). Panel ({\bf b}): $|\eta_2|/\Gamma=5\times 10^{-4}$ (black
  curve), $|\eta_2|/\Gamma=5\times 10^{-5}$ (red curve),
  $|\eta_2|/\Gamma=2\times 10^{-5}$ (blue curve), $|\eta_2|/\Gamma=10^{-5}$
  (green curve). The other parameters are the same as in
  Fig. \ref{figure_2}.}
\end{figure}
$\partial S^>(\omega,V,\Delta\phi)/\partial V$ characterized by the universal
unitary maximum $3e^3/4h$ at the frequency $\hbar\omega=|eV|/2$ is a unique
signature that the BIC has the Majorana origin and does not result from other
quantum states. On the other side, since mean currents might be controversial
with respect to Majorana states, the non-Majorana nature of the BIC could have
still been assumed if one had restricted the analysis of the BIC only by the
differential conductance measurements \cite{Ramos-Andrade_2019} which are
still important and should be performed as the first step before probing the
BIC via more involved measurements such as, for example, measurements of the
differential quantum noise at finite frequencies. We emphasize that the the
Majorana interference plays here a fundamental role activating the
photon-absorption channel around $\hbar\omega=|eV|/2$ for low bias voltages
$|eV|\ll\Gamma$. Indeed, as we have already seen in Fig. \ref{figure_2},
although opening of this photon-absorption channel is also admitted for
$|\eta_2|=0$, without the Majorana interference this channel is not active in
the sense that it does not produce any additional finite contribution to the
differential quantum noise $\partial S^>(\omega,V,\Delta\phi)/\partial V$.

To see how the universal resonance develops and disappears at low bias
voltages, $|eV|\ll\Gamma$, we analyze the value of the differential quantum
noise at $\hbar\omega=|eV|/2$ as a function of $\Delta\phi$ for various values
of $|\eta_2|$. From the previous discussion we know that, on one side, this
resonance must disappear for $\Delta\phi\rightarrow 0,\pi$ or
$|\eta_2|\rightarrow 0$ and that, on the other side, its universal unitary
maximum $3e^3/4h$ is already reached for very small values of the Majorana
phase difference, such as $\Delta\phi=\pi/128$, when $|\eta_2|$ is
finite. Thus one could naturally conclude that for a finite value of
$|\eta_2|$ the resonance arises as a jump for arbitrarily small deviations of
$\Delta\phi$ from $0$ and $\pi$. However, our numerical results shown in
Fig. \ref{figure_4} demonstrate that this is not so and the behavior of the
resonance is highly nontrivial. As one can see in Fig. \ref{figure_4}(a)
showing results for larger values of $|\eta_2|$, the differential quantum
noise at $\hbar\omega=|eV|/2$ indeed quickly grows from the universal unitary
plateau $e^3/4h$ at $\Delta\phi=0,\pi$ to the universal unitary maximum
$3e^3/4h$. However, if $|\eta_2|$ is very large,
$\partial S^>(\omega,V,\Delta\phi)/\partial V$ is strongly suppressed far from
$\Delta\phi=0,\pi$ as demonstrated in Fig. \ref{figure_4}(a) by the black
curve corresponding to the value of $|\eta_2|$ which has also been used to
obtain the black and red curves in Fig. \ref{figure_3}. Thus for large values
of $|\eta_2|$ the resonance is fully developed in small regions centered
around the points $\Delta\phi=0,\pi$ in whose extremely narrow vicinities
(located fully within these small regions) the resonance disappears recovering
the universal unitary plateau $e^3/4h$. The inset in Fig. \ref{figure_4}(a)
shows one of these extremely narrow vicinities, namely, the one with
$\Delta\phi=\pi$. Clearly, these extremely narrow vicinities should expand
when $|\eta_2|$ decreases and this is what one observes in the inset. Upon
decreasing $|\eta_2|$ such expanding should eventually decrease the
differential quantum noise in the small regions, where
$\partial S^>(\omega,V,\Delta\phi)/\partial V=3e^3/4h$, and increase
$\partial S^>(\omega,V,\Delta\phi)/\partial V$ in the wide regions, where it
is strongly suppressed, making it equal to the universal unitary plateau
$e^3/4h$ in the whole range of $\Delta\phi$ for sufficiently small values of
$|\eta_2|$. However, before this situation is achieved, one observes that when
$|\eta_2|$ decreases, the resonance is fully developed in wider and wider
regions of the Majorana phase difference as it is demonstrated in
Fig. \ref{figure_4}(b). Indeed, as one can see, the black and red curves in
Fig. \ref{figure_4}(b) have wide plateaus on which
$\partial S^>(\omega,V,\Delta\phi)/\partial V=3e^3/4h$. This corresponds to
the situation when the resonance fully develops already for very small
deviations from $\Delta\phi=0,\pi$ and reaches its universal unitary maximum
$3e^3/4h$ almost in the whole range of $\Delta\phi$. Note that this behavior
happens when $|\eta_2|$ is about seven orders of magnitude smaller than
$|\eta_1|$. This emphasizes that even very weak coupling of the second
Majorana mode $\gamma_2$ to the QD may fundamentally change the behavior of
the differential quantum noise. Only for very small values of $|\eta_2|$ the
universal unitary plateaus $3e^3/4h$ are destroyed and, as can be seen from
the blue and green curves in Fig. \ref{figure_4}(b),
$\partial S^>(\omega,V,\Delta\phi)/\partial V$ starts to decrease down to the
universal unitary plateau $e^3/4h$ for all values of $\Delta\phi$.
\begin{figure}
\includegraphics[width=8.0 cm]{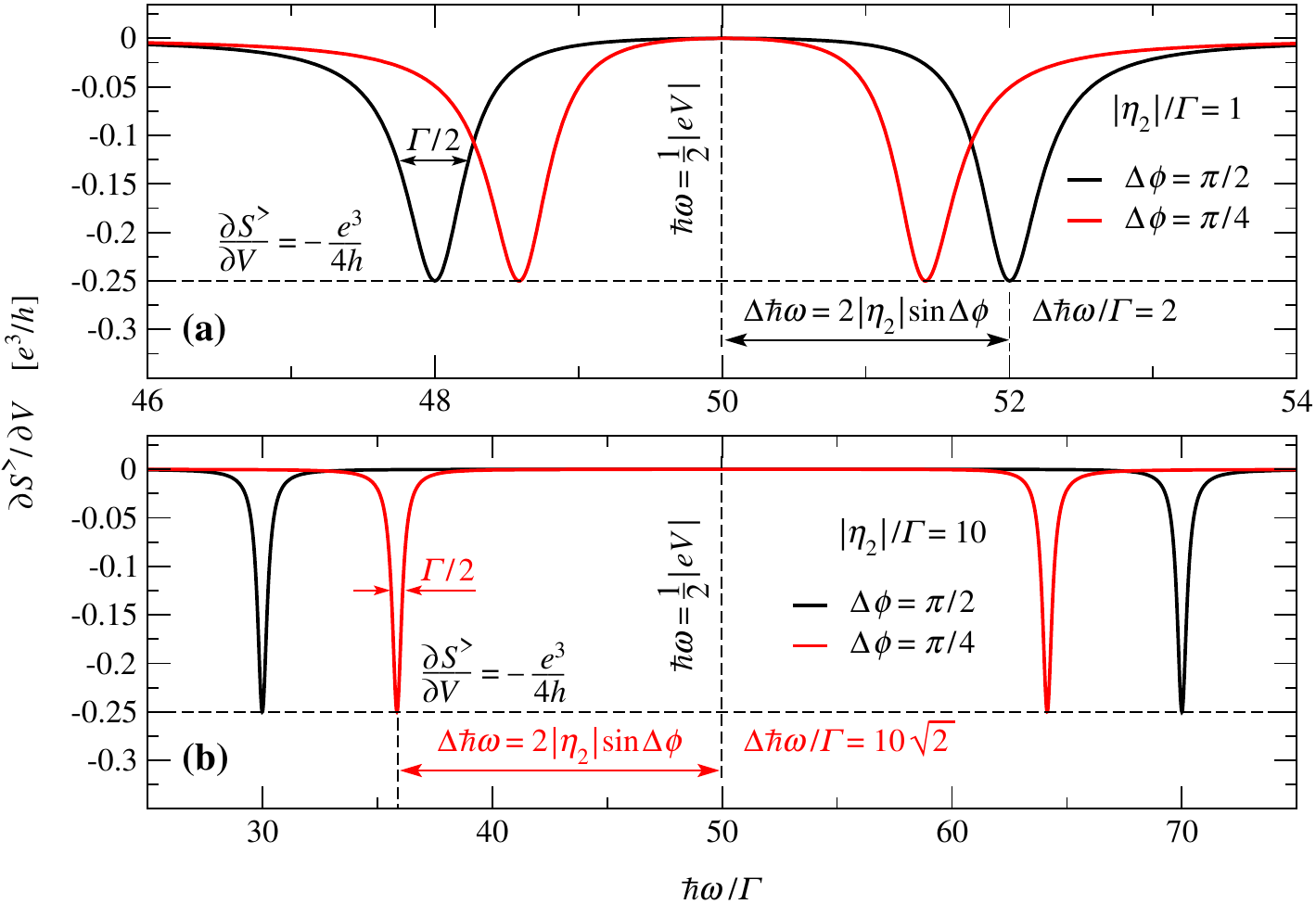}
\caption{\label{figure_5} Differential quantum noise
  $\partial S^>(\omega,V,\Delta\phi)/\partial V$ as a function of the
  frequency $\omega$ at high bias voltages, $|eV|\gg\Gamma$, and in the
  presence of direct tunneling between the QD and the second Majorana mode
  $\gamma_2$. For all the curves $|eV|/\Gamma=10^2$. Panel ({\bf a}):
  $|\eta_2|/\Gamma=1$ and $\Delta\phi=\pi/2$ (black curve), $\Delta\phi=\pi/4$
  (red curve). Panel ({\bf b}): $|\eta_2|/\Gamma=10$ and $\Delta\phi=\pi/2$
  (black curve), $\Delta\phi=\pi/4$ (red curve). The other parameters are the
  same as in Fig. \ref{figure_2}.}
\end{figure}

Majorana quantum noise in strongly nonequilibrium states induced by high bias
voltages, $|eV|\gg\Gamma$, is particularly appealing for future
experiments. Indeed, in this case
$\partial S^>(\omega,V,\Delta\phi)/\partial V$ turns out to be extremely
sensitive to the Majorana interference especially for
$|\eta_2|\gtrsim\Gamma$. As has been shown in Fig. \ref{figure_2}, for
$|\eta_2|=0$ and high bias voltages, $|eV|\gg\Gamma$, the differential quantum
noise is suppressed for all frequencies except for a vicinity of
$\hbar\omega=|eV|/2$. Around this frequency there arises an antiresonance with
the universal unitary minimum $-e^3/4h$ located at $\hbar\omega=|eV|/2$. The
full width of this antiresonance at half of its minimum is equal to
$\Gamma$. As demonstrated by Fig. \ref{figure_5}, in the presence of direct
tunneling between the QD and the second Majorana mode $\gamma_2$, that is when
$|\eta_2|\neq 0$, this antiresonance is split into two antiresonances. From
our numerical calculations we find that the minima of these two antiresonances
are located symmetrically with respect to the frequency $\hbar\omega=|eV|/2$,
that is equidistantly from the location of the minimum of the original
antiresonance, specifically, at the two frequencies
\begin{equation}
  \hbar\omega_\pm=\frac{|eV|}{2}\pm 2|\eta_2|\sin\Delta\phi.
  \label{Antres_f_pm_hbv}
\end{equation}
At these two frequencies the differential quantum noise reaches the same
universal unitary minimum $-e^3/4h$ as the one of the original
antiresonance. Moreover, the two antiresonances centered around
$\hbar\omega_\pm$ are twice narrower than the original antiresonance from
which they have emerged via the above mentioned splitting. That is they have
the same full widths $\Delta\hbar\omega$, equal to $\Gamma/2$, at half of
their universal unitary minimum $-e^3/4h$,
\begin{equation}
  \begin{split}
    &\frac{\partial S^>(\omega,V,\Delta\phi)}{\partial V}\biggl|_{\omega=\omega_\pm}=-\frac{e^3}{4h},\\
    &\Delta\hbar\omega=\frac{\Gamma}{2}.
  \end{split}
  \label{Uumin_f_pm_hbv}
\end{equation}
Alternatively, instead of the above described picture illustrating the
emergence of the above two antiresonances as a split of the original
antiresonance one might conceive of a resonance induced by the Majorana
interference and interpret it in terms of the BIC discussed in connection with
Fig. \ref{figure_3}. This resonance develops in the middle of the original
antiresonance as soon as $|\eta_2|\neq 0$. The height of this resonance
\begin{figure}
\includegraphics[width=8.0 cm]{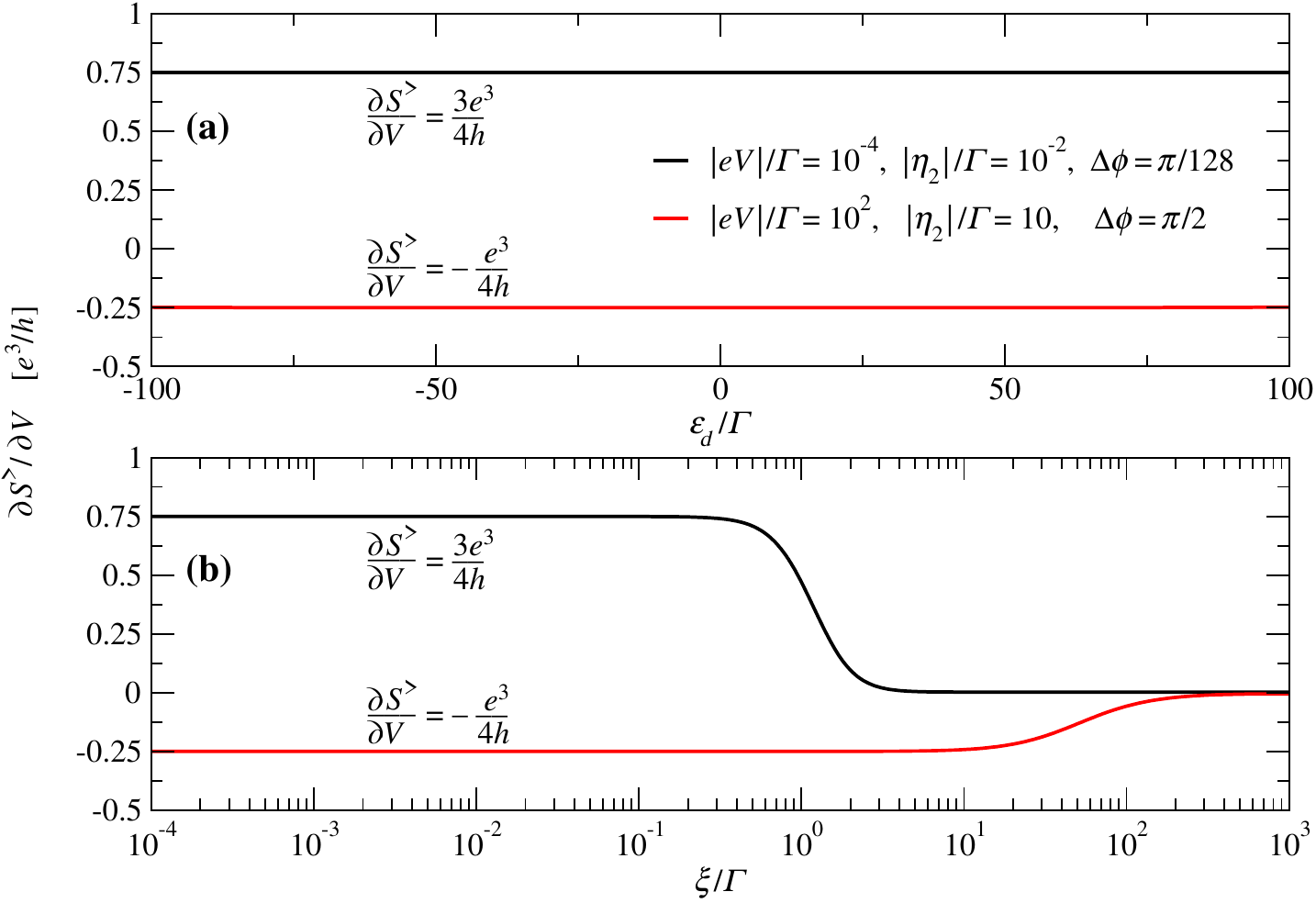}
\caption{\label{figure_6} Differential quantum noise
  $\partial S^>(\omega,V,\Delta\phi)/\partial V$ at the resonance and
  antiresonance frequencies. Panel ({\bf a}): As a function of the QD energy
  level $\epsilon_d$ at the resonance frequency $\hbar\omega=|eV|/2$ (black
  curve) and antiresonance frequencies $\omega=\omega_\pm$ (red curve). For
  both curves the overlap energy $\xi/\Gamma=10^{-14}$. Panel ({\bf b}): As a
  function of the overlap energy $\xi$ at the resonance frequency
  $\hbar\omega=|eV|/2$ (black curve) and antiresonance frequencies
  $\omega=\omega_\pm$ (red curve). For both curves the QD energy level
  $\epsilon_d/\Gamma=10$. In both panels for the black curves
  $|eV|/\Gamma=10^{-4}$, $|\eta_2|/\Gamma=10^{-2}$, $\Delta\phi=\pi/128$
  whereas for the red curves $|eV|/\Gamma=10^2$, $|\eta_2|/\Gamma=10$,
  $\Delta\phi=\pi/2$. The other parameters are the same as in
  Fig. \ref{figure_2}.}
\end{figure}
approaches the universal unitary value $e^3/4h$ when its maximal value
approaches zero at $\hbar\omega=|eV|/2$. The bottom of the resonance is
reached at the two points $\hbar\omega_\pm$ with the distance between them
\begin{equation}
  |\hbar\omega_+-\hbar\omega_-|=4|\eta_2||\sin\Delta\phi|.
  \label{Dist_f_pm_hbv}
\end{equation}
In experiments one may vary the Majorana phase difference $\Delta\phi$ and
observe how the two antiresonances move with respect to each other. Measuring
then the maximal distance between the minima of the antiresonances one obtains
the maximal value of $|\hbar\omega_+-\hbar\omega_-|$, that is
$4|\eta_2|$. After that for any given distance between the minima of the
antiresonances, $|\hbar\omega_+-\hbar\omega_-|$, one may obtain the
corresponding Majorana phase difference as
$|\sin\Delta\phi|=|\hbar\omega_+-\hbar\omega_-|/4|\eta_2|$. Therefore
detecting the two antiresonances, specifically, the locations of their minima
at high bias voltages, $|eV|\gg\Gamma$, would enable to determine the
frequencies $\hbar\omega_\pm$ and, in this way, measure the values of
$|\eta_2|$ and $\Delta\phi$ in experiments on nonequilibrium finite frequency
quantum noise driven by Majorana interference.

Let us now verify the universality of the Majorana resonance and
antiresonances at, respectively, low and high bias voltages. The universality
assumes independence of the resonance and antiresonances on the gate voltage
controlling the position of the QD energy level $\epsilon_d$. As one can see
in Fig. \ref{figure_6}(a), both the resonance and antiresonances remain
unchanged when $\epsilon_d$ is varied over a wide range. Note, that here we
have also included negative values of $\epsilon_d$ to confirm our above
statement that in the universal Majorana regime, Eq. (\ref{U_M_regime}), the
differential quantum noise does not depend on $\epsilon_d$ both for
$\epsilon_d>0$ and $\epsilon_d<0$.
\begin{figure}
\includegraphics[width=8.0 cm]{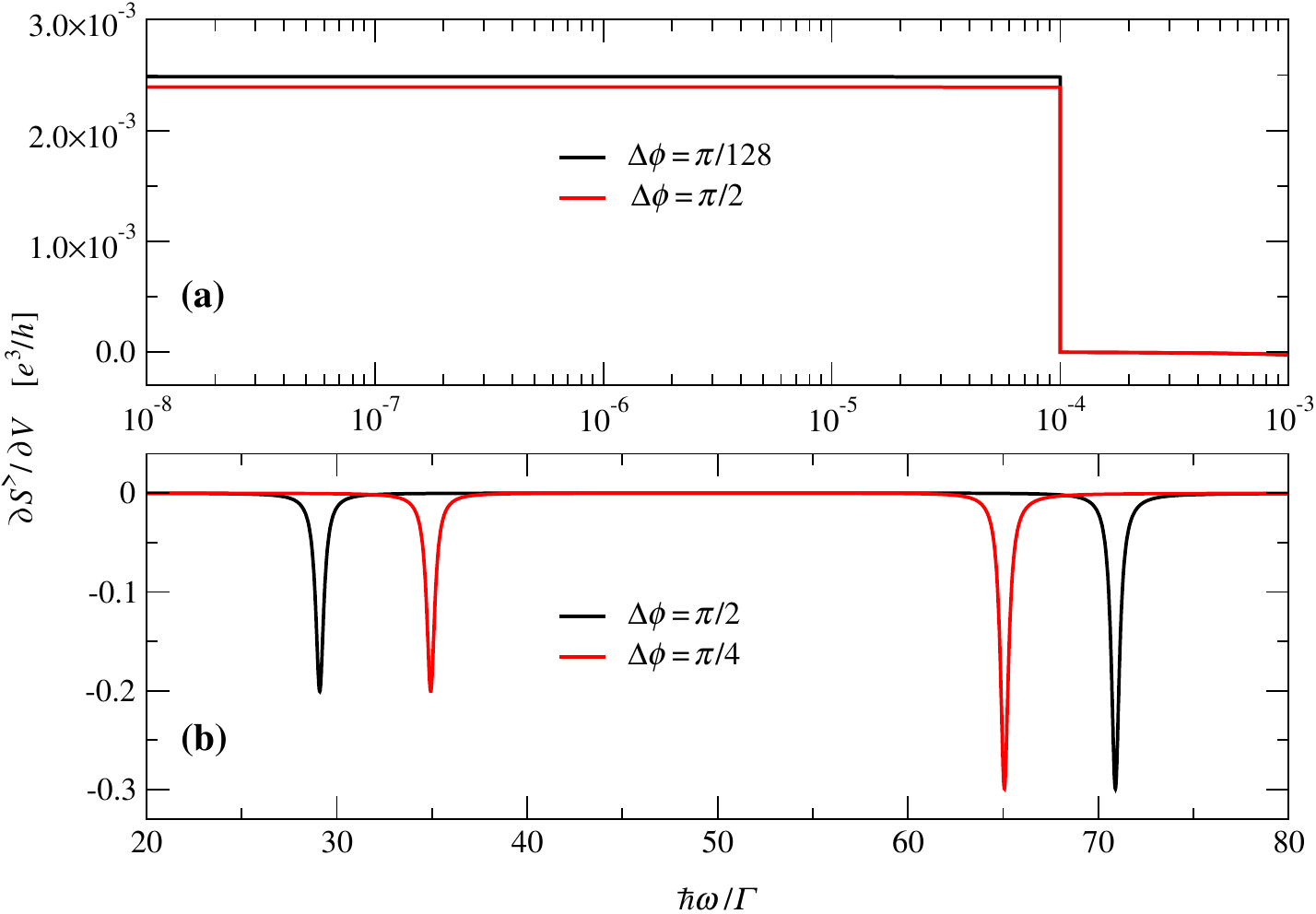}
\caption{\label{figure_7} Differential quantum noise
  $\partial S^>(\omega,V,\Delta\phi)/\partial V$ as a function of the
  frequency $\omega$ for strongly overlapping MBSs characterized by
  $\xi/\Gamma=2\times 10^2$. Panel ({\bf a}): Low bias voltages,
  $|eV|\ll\Gamma$, with the specific values of the parameters
  $|\eta_2|/\Gamma=10^{-2}$, $|eV|/\Gamma=10^{-4}$ and $\Delta\phi=\pi/128$
  (black curve), $\Delta\phi=\pi/2$ (red curve). Panel ({\bf b}): High bias
  voltages, $|eV|\gg\Gamma$, with the specific values of the parameters
  $|\eta_2|/\Gamma=10$, $|eV|/\Gamma=10^2$ and $\Delta\phi=\pi/2$ (black
  curve), $\Delta\phi=\pi/4$ (red curve). The other parameters are the same as
  in Fig. \ref{figure_2}.}
\end{figure}

It is also important to explore how the differential quantum noise at the
resonance and antiresonance frequencies behaves at large values of $\xi$ that
is when the overlap of the MBSs is strong as may happen for short distances
between the MBSs. In this case the MBSs communicate through the TS and this
communication may provide an additional interference
channel. Fig. \ref{figure_6}(b) demonstrates that when the MBSs start to
strongly overlap, the differential quantum noise is significantly suppressed
both at the resonance frequency $\hbar\omega=|eV|/2$ for low bias voltages,
$|eV|\ll\Gamma$, and at the antiresonance frequencies $\omega_{\pm}$ for high
bias voltages, $|eV|\gg\Gamma$. To explore in more detail the fate of the
resonance and antiresonances for strongly overlapping MBSs we have analyzed
the frequency dependence of the differential quantum noise at large values of
$\xi$. In the regime of low bias voltages, $|eV|\ll\Gamma$, shown in
Fig. \ref{figure_7}(a), we find that the communication between the MBSs
through the TS results in a disappearance of the resonance at
$\hbar\omega=|eV|/2$ for all values of the Majorana phase difference
$\Delta\phi$. Moreover, the dependence on $\Delta\phi$ becomes very weak as
demonstrated by the black and red curves which are very close to each other
even for the values of $\Delta\phi$ chosen to reach the maximal distance
between the curves. Although the step-like behavior (with the jump located at
$\hbar\omega=|eV|$) is still present, the differential quantum noise is
significantly suppressed below the universal unitary plateau arising at small
values of $\xi$, that is
$\partial S^>(\omega,V,\Delta\phi)/\partial V\ll e^3/4h$. A more interesting
behavior is observed in the regime of high bias voltages, $|eV|\gg\Gamma$,
shown in Fig. \ref{figure_7}(b). Here we find two major changes in comparison
with the picture discussed above for small values of $\xi$. First, the minima
of the two antiresonances become unequal and significantly deviate from the
universal unitary minimum $-e^3/4h$. Second, the locations of these minima
shift from the frequencies $\omega_\pm$ (see Eq. (\ref{Antres_f_pm_hbv})) to
some other frequencies. For larger values of $\xi$ both the deviations from
the universal unitary minimum and the shifts from the frequencies $\omega_\pm$
increase even further.

Finally, we would like to compare the above discussed behavior of the
differential quantum noise induced by the MBSs, in particular its resonance
and antiresonances brought by the Majorana interference, with the differential
quantum noise induced by ABSs. Transport phenomena relevant to ABSs are
captured by our theoretical model in a certain range of its
parameters. Specifically, when the overlap energy $\xi$ is large and the
Majorana tunneling amplitudes are of the same order, $|\eta_2|\sim |\eta_1|$,
our theoretical model describes (see Ref. \cite{Deng_2018}) ABSs coupled to
the QD. Note, that a different regime which is also specified by the Majorana
tunneling amplitudes of the same order, $|\eta_2|\sim |\eta_1|$, but in which
the overlap energy $\xi$ remains small would not be appropriate to model ABSs
because it would describe the MBSs in a system with a curved TS corresponding
to Fig. \ref{figure_1}(b) whose analysis we would like to postpone for a
future research. To understand how the differential quantum noise changes when
the MBSs are replaced with the ABSs we have performed a numerical analysis for
the case of ABSs coupled to the QD. It turns out that, in contrast to the
MBSs, the differential quantum noise driven by the ABSs does not have any
resonance at the frequency $\hbar\omega=|eV|/2$ in the regime of low bias
voltages, $|eV|\ll\Gamma$. Indeed, as demonstrated in Fig. \ref{figure_8}(a),
the differential quantum noise has the expected step-like behavior (with the
jump located at $\hbar\omega=|eV|$) but no any resonance at
$\hbar\omega=|eV|/2$ appears for any value of $\Delta\phi$. Moreover, the
differential quantum noise is strongly suppressed below the universal unitary
plateau $e^3/4h$. We find that this suppression cannot be compensated by
varying $\Delta\phi$ whose increase suppresses
$\partial S^>(\omega,V,\Delta\phi)/\partial V$ even further as demonstrated by
the red curve. In the regime of high bias voltages, $|eV|\gg\Gamma$, the ABSs
have much more interesting fluctuation fingerprints which are essentially
different from the ones characterizing the MBSs. Recall that the MBSs give
rise to two antiresonances with equal universal unitary minima at which
$\partial S^>(\omega,V,\Delta\phi)/\partial V=-e^3/4h$. The minima are located
at the two frequencies $\omega_\pm$ (see Fig. \ref{figure_5} and
Eq. (\ref{Antres_f_pm_hbv})). The distance between these antiresonances
depends on both $|\eta_2|$ and $\Delta\phi$. In contrast, as one can see in
Fig. \ref{figure_8}(b), the ABSs give rise to resonance and antiresonance and
not to a pair of antiresonances as it happens for the MBSs. Further, in these
resonance-antiresonance pairs induced by the ABSs the amplitudes of the
resonance and antiresonance are not equal. Moreover, the frequencies of the
ABS resonance and antiresonance are different from the frequencies
$\omega_\pm$ of the MBS antiresonances. As demonstrated by the black and red
curves in Fig. \ref{figure_8}(b), the frequencies of the ABS resonance and
antiresonance depend on $\Delta\phi$. They also depend on $|\eta_2|$ as
demonstrated by the black and green curves. However, in contrast to the
distance between the MBS antiresonances, the distance between the ABS
resonance and antiresonance depends neither on $\Delta\phi$ nor on $|\eta_2|$
as one clearly observes from the black, red and green curves. Our numerical
analysis reveals that in the ABS range of our model and in the regime
specified by Eq. (\ref{U_M_regime}) the distance between the ABS resonance and
antiresonance depends only on the bias voltage and, in fact, it is equal to
$|eV|$ as, in particular, exemplified by the green and blue curves.

\section{Conclusion}\label{concl}
In this work we have numerically investigated universal fluctuation
fingerprints of Majorana interference in the differential quantum noise
$\partial S^>(\omega,V,\Delta\phi)/\partial V$ at finite frequencies $\omega$
in a system where a QD is linked via tunneling to MBSs of a TS with the
tunneling phase difference $\Delta\phi$. Nonequilibrium states of this system
are induced by a bias voltage $V$. Both low and high bias voltages have been
considered. In the regime of low bias voltages it has been found that when the
MBSs do not interfere, the differential quantum noise as a function of the
frequency has a step-like shape with the universal unitary Majorana plateau
$e^3/4h$ for $\hbar\omega<|eV|$ and with strong suppression of
$\partial S^>(\omega,V)/\partial V$ for $\hbar\omega>|eV|$. In presence of the
Majorana interference we have discovered that in a vicinity of the frequency
$\hbar\omega=|eV|/2$ there develops a narrow resonance whose characteristic
width is proportional to $\sin^2\Delta\phi$. The maximum of this resonance
reaches another universal unitary Majorana value, namely $3e^3/4h$, at
$\hbar\omega=|eV|/2$. The appearance of this resonance has been explained in
terms of an additional photon-absorption channel. This channel is
energetically admitted by weak excitations of the QD and tunneling processes
from the left contact to the TS. However, opening of this additional
photon-absorption channel is activated only by the Majorana interference which
gives an additional finite contribution in the form of the above Majorana
resonance in $\partial S^>(\omega,V,\Delta\phi)/\partial V$. Exploring the
regime of high bias voltages we have found that in absence of the Majorana
interference the differential quantum noise as a function of the frequency is
strongly suppressed everywhere except for a vicinity of the frequency
$\hbar\omega=|eV|/2$. Here $\partial S^>(\omega,V)/\partial V$ has an
antiresonance with the minimum equal to the universal unitary Majorana value
$-e^3/4h$. This minimum is located at $\hbar\omega=|eV|/2$. The characteristic
width of this antiresonance is determined by the strength of the tunneling
between the QD and contacts. When the Majorana interference appears, this
antiresonance splits into two antiresonances. The minima of these
antiresonances are the same as the minimum of the original antiresonance, that
is they both reach the universal unitary Majorana value $-e^3/4h$. They are
located at two finite frequencies $\hbar\omega_\pm$ specified by $V$ and
$\Delta\phi$. The characteristic widths of the two antiresonances are equal
and twice less than the width of the original antiresonance. An alternative
interpretation of the two antiresonances has been given in terms of a
resonance emerging at the minimum of the original antiresonance. Specifically,
for high bias voltages the Majorana interference also activates the additional
photon-absorption channel discussed for the case of low bias voltages and
gives an additional finite contribution to the differential quantum
noise. This finite contribution forms in
$\partial S^>(\omega,V,\Delta\phi)/\partial V$ a resonance at the minimum of
the original antiresonance and, as a result, there appear two antiresonances
at the frequencies $\hbar\omega_\pm$. It has been suggested that a detection
of the frequencies $\hbar\omega_\pm$ might be a practical tool to measure the
Majorana phase difference $\Delta\phi$ in experiments on nonequilibrium finite
frequency quantum noise. In addition, the universality of the Majorana
resonance and antiresonances at, respectively, low and high bias voltages has
been demonstrated and the case of strongly overlapping MBSs has been
analyzed. Finally, the fluctuation fingerprints explored for the MBSs have
been compared with those emerging when the MBSs are replaced with ABSs coupled
to the QD. It has been shown that the ABSs give rise to fundamental changes in
the behavior of the differential quantum noise as compared to the one induced
by the MBSs. In particular, for the ABSs one does not observe any resonance at
$\hbar\omega=|eV|/2$ in the low bias regime whereas in the high bias regime
instead of two antiresonances with equal amplitudes there appear a resonance
and antiresonance with different amplitudes. The frequencies of these ABS
resonance and antiresonance are different from the frequencies of the two MBS
antiresonances. Moreover, in contrast to the distance between the MBS
antiresonances, the distance between the ABS resonance and antiresonance does
not depend on $|\eta_2|$ and $\Delta\phi$ but depends only on the bias
voltage. Specifically, our numerical analysis has revealed that this distance
is equal to $|eV|$.

In the present work we have focused on numerical calculations of the
differential quantum noise $\partial S^>(\omega,V,\Delta\phi)/\partial V$ at
finite frequencies $\omega$ and predicted an appearance of a resonance and
antiresonances induced by Majorana interference. The numerical approach has
allowed us to achieve the goals formulated in the introduction, Section
\ref{intro}, in particular, to reveal universal unitary values characterizing
fluctuation fingerprints of interfering MBSs at finite frequencies
$\omega$. Namely, the maximum and minima of, respectively, the Majorana
resonance and antiresonances are quantized to some specific fractions of
$e^3/h$. Therefore, there appears a natural question: How should one
understand these fractions? Within a pure numerical approach it is hard or may
be even impossible to answer this interesting question. Thus, as a possible
outlook, it would be useful to obtain an analytical solution of the problem or
develop an appropriate effective model which would be able to explicitly
demonstrate how the interfering MBSs result in the numerically predicted
fractional values of the differential quantum noise
$\partial S^>(\omega,V,\Delta\phi)/\partial V$ at finite frequencies $\omega$
in nonequilibrium states induced by low and high bias voltages $V$. Another
interesting problem is to investigate universal fingerprints of Majorana
interference in the finite frequency quantum noise when nonequilibrium states
of a system with MBSs have thermoelectric nature. This might be achieved, {\it
  e.g.}, by applying to the system both electric and thermal voltages. If the
MBSs interfere, one may expect that the differential quantum noise will have a
specific universal behavior at finite frequencies. How these frequencies
depend on various parameters controlling both the Majorana interference and
nonequilibrium in the system is an important problem for fundamental and
practical research on MBSs.
\section*{Acknowledgments}
The author sincerely thanks Reinhold Egger for a useful discussion of the
results presented in the paper.

\end{document}